\documentclass[preprint2]{aastex}

\usepackage{amsmath,subfigure,epsfig,natbib,mathptmx}
\slugcomment{group size, relatedness, kin selection, multi-level selection, linkage disequilibrium, Snowdrift game.}

\begin{document}
\title{The Concurrent Evolution of Cooperation and the Population Structures that Support it}
\author{Simon T. Powers, Alexandra S. Penn \and Richard A. Watson}
\affil{Science \& Engineering of Natural Systems Group, School of Electronics \& Computer Science, University of Southampton, Southampton, SO17 1BJ, U.K.}
\email{Simon.Powers@Unil.ch; asp@ecs.soton.ac.uk; raw@ecs.soton.ac.uk}

\begin{abstract}

The evolution of cooperation often depends upon population structure, yet nearly all models of cooperation implicitly assume that this structure remains static. This is a simplifying assumption, since most organisms possess genetic traits that affect their population structure to some degree. These traits, such as a group size preference, affect the relatedness of interacting individuals and hence the opportunity for kin or group selection. We argue that models which do not explicitly consider their evolution cannot provide a satisfactory account of the origin of cooperation, since they cannot explain how the prerequisite population structures arise. Here we consider the concurrent evolution of genetic traits that affect population structure, with those that affect social behaviour. We show that not only does population structure drive social evolution, as in previous models, but that the opportunity for cooperation can in turn drive the creation of population structures that support it. This occurs through the generation of linkage disequilibrium between socio-behavioural and population-structuring traits, such that direct kin selection on social behaviour creates indirect selection pressure on population structure. We illustrate our argument with a model of the concurrent evolution of group size preference and social behaviour. [This is a post-print of an accepted manuscript published in \textit{Evolution} 65(6) pp. 1527-1543, June 2011. The publisher's version is available from http://onlinelibrary.wiley.com/doi/10.1111/j.1558-5646.2011.01250.x/full.]

\end{abstract}

It is widely appreciated that population structure drives the evolution of social traits. Cooperative behaviours, that benefit other individuals at some cost to the actor, can evolve if the population structure is such that the benefits of cooperation fall upon other cooperators, rather than being distributed uniformly \citep{Hamilton:1964:a,Michod:1985:a,Lehmann:2006:a,Fletcher:2009:a}. Over the past 50 years, social evolution theory has focussed on determining the range of population structures that can support cooperation (e.g., \citealt{Smith:1964:a,Levin:1974:a,Wilson:1975:a,Wilson:1987:a,Nowak:1992:a,Mitteldorf:2000:a,Hauert:2004:a,Peck:2004:a,elMouden:2010:a}). For example, cooperation might be selectively favoured if organisms live in groups below a certain size, and selfish behaviour favoured in larger groups \citep{Levin:1974:a,Wilson:1981:a}. Crucially, such models assume that population structure is not directly affected by individual traits (although some models do allow for an elastic population structure, such that the average level of cooperation within a group affects group size \citep{Grafen:2007:a,Wild:2009:a} or dispersal probability \citep{Pepper:2002:a}). This means that cooperation can then simply be viewed as the adaptation of organisms' social behaviour to the social environment that they find themselves in.  What such models cannot explain, however, is why organisms live in a population structure that supports a certain level of cooperation. We argue that answering this question is necessary in order to explain the evolutionary origin of cooperation, for a satisfactory account of the origin of cooperation must explain why a social environment that selects for cooperative interactions exists.

Population structure is the product not only of environmental factors, but also of individual behaviours. Many of these behaviours that affect population structure have a genetic basis, and so are themselves subject to natural selection. For example, the evolution of individual traits that affect group size \citep{Rodman:1981:a,Koenig:1981:a,Giraldeau:2000:a,Kokko:2001:a}, dispersal rate \citep{Johnson:1990:a}, or aspects of the mating system \citep{Orians:1969:a,Emlen:1977:a} have all been considered in the literature.  Despite this, however, there have been very few treatments that consider how their evolution is affected by selection pressures on (individually-costly) cooperative behaviour. Here, we generalise the small body of literature that has considered particular cases in which social behaviour co-evolves with the mating system \citep{Peck:1988:a,Breden:1991:a}, group size \citep{Aviles:2002:a}, or dispersal rate \citep{LeGalliard:2005:a,Hochberg:2008:a}. Specifically, we present here a general logical argument which demonstrates that the benefits of cooperation can drive the evolution of population structure, leading to the creation of population structures that support cooperative behaviour. We then illustrate our general argument with a numerical model that considers the concurrent evolution of cooperation with a particular population-structuring trait; specifically, a preference for the number of individuals that found a group (which we refer to as founding size). Using this model, we are able to show conditions under which organisms can evolve from living in a population structure that supports little cooperation (i.e. large interaction groups, being approximately equivalent to no population structure, where selfish behaviour dominates), to one where cooperative behaviour predominates. We thus consider a population structure that initially approaches freely-mixed conditions, where individuals have fitness-affecting interactions with many others, and then show how greater interaction structure can evolve. Specifically, this illustrates how evolution of an individual group size preference can increase between-group variance / genetic relatedness (group and kin selection are equivalent ways of understanding the selection pressures that population structure exerts on social traits \citep{Hamilton:1975:a,Queller:1992:a,Foster:2006:a,Foster:2006:c,Lehmann:2007:a}; see Discussion section).

Our approach is in contrast to recent work by \citet{Aviles:2002:a}, which shows how solitaires can evolve to live in groups (i.e., how a starting group size of 1 can evolve upwards). This fundamental difference is explained by the type of cooperative act that the respective models seek to explain. Specifically, Avil\'{e}s considers facultative types of social interaction, i.e., some task that it is possible to do alone but which can be done more efficiently in a group \citep{vanVeelen:2010:a}. By contrast, we consider obligatory social interactions, where it is a fact of life that organisms must interact with others (but not that the interaction is cooperative). For example, it is usually not possible for an organism to be a solitaire with respect to common resources, and resource usage rate can create a social dilemma with cooperative and selfish strategies \citep{Kreft:2004:a,Zea-Cabrera:2006:a}. Likewise, a micro-organism may need to produce extra-cellular goods to survive \citep{Ajit:2007:a,Gore:2009:a}; the fact that they are extracellular means that they are shared with others, automatically creating a social trait-group structure \citep{Wilson:1975:a,Wilson:1980:a}. In such cases of obligatory social interactions, where organisms cannot be solitaires, the evolution of population structure means the refinement of an always present trait-group structure. A group size of one would therefore be an invalid starting assumption for such interactions. Our approach to the evolution of group size is also fundamentally different to other recent work in this area, such as models of optimal group size in the context of social foraging \citep{Giraldeau:2000:a}, or group augmentation during cooperative breeding \citep{Kokko:2001:a}. These works consider how direct fitness benefits (for example, due to an Allee effect increasing individual fitness with group size; \citealt{Allee:1938:a,Aviles:1999:a}) of being in a group can drive group size upwards. By contrast, we focus on indirect benefits that arise through increased kin or group selection for greater (individually-costly) cooperation. That is, we show how the benefits of cooperation can drive the evolution of structures that increase relatedness, and hence create the conditions for effective kin selection.  

The kind of social interactions that we consider can be readily modelled using the Prisoner's Dilemma and Snowdrift games. Such evolutionary game theoretic models are commonly used to conceptualise problems of cooperation across taxa \citep{Smith:1982:a}, including in the literature on cooperation in animals, humans, and microbes \citep{Grafen:1979:a,Dugatkin:1990:a,Nowak:1992:a,Frick:2003:a,Greig:2004:a,Doebeli:2005:a,Kun:2006:a,Kummerli:2007:a,Gore:2009:a,Stark:2010:a}, and can handle interactions in groups of size $n$ as well as between pairs of individuals. In particular, the $n$-player Prisoner's Dilemma provides a simple way to model directional selection within social groups favouring selfish behaviour (and is the implicit assumption in most models of social evolution \citep{Fletcher:2007:a}), whilst the Snowdrift game provides a simple model of negative frequency-dependent selection leading to a polymorphism of selfish cheats and cooperative individuals within a group \citep{Doebeli:2005:a}. Examples of where selection on social behaviour may be negative frequency-dependent in this way can be found across taxa and include enzyme secretion in yeast \citep{Gore:2009:a} and viruses, antibiotic resistance in bacteria \citep{Dugatkin:2003:a,Dugatkin:2005:a}, social foraging in spider colonies \citep{Pruitt:2009:a}, and sentinel behaviour in mammals (see also \citealt{Doebeli:2005:a} for a review). Our model illustrates that the difference between directional and frequency-dependent selection on social behaviour has a profound effect on the evolution of population structure. Specifically, evolution of population structure in the direction that increases cooperation can occur from a much larger range of conditions where social interactions are of the Snowdrift type, that is, where selection supports a polymorphism of social behaviours within a group.

\section*{The benefits of cooperation can drive the evolution of population structure}

\citet{Peck:1988:a} and \citet{Breden:1991:a} argue that cooperation can drive the evolution of population structure, for the specific case of evolution of aspects of the mating system. Here, we develop a general argument that applies in principle to any heritable trait that affects an aspect of the population structure of its bearers. On the one hand, it may seem obvious that a population structure which supports cooperation would be preferred over one which does not, since cooperation by definition raises absolute individual fitness. On the other hand, individual selection responds to relative and not absolute fitness, and selfish individuals by definition have a greater relative fitness than those cooperators which they exploit. Therefore, one might expect that selfish individuals will prevail due to their relative fitness advantage, and create population structures which support themselves rather than cooperators. Accordingly, the result of concurrent evolution between social traits and population-structuring traits is not obvious. But, below we argue that linkage disequilibrium will be generated between these two types of traits, and that this allows the mean fitness advantage of cooperation to be realised.

Consider two possible population structures, such as two different group sizes or two different mating systems. Suppose that structure $A$ causes selection to favour more cooperative behaviour amongst its inhabitants, while structure $a$ causes selection for more selfish behaviour. For example, structure $A$ might be a smaller initial group size \citep{Levin:1974:a,Wilson:1981:a}, or a greater degree of inbreeding \citep{Wade:1981:a}. Furthermore, consider organisms that posses two heritable traits, the first of which controls their social behaviour (cooperative or selfish), and the second controls the population structure that they live in ($A$ or $a$). If the two traits start off in linkage equilibrium, then linkage disequilibrium between them will evolve, by the following logic. The more cooperative allele will increase in frequency more in structure $A$ than it does in structure $a$. Moreover, the individuals that live in structure $A$ are those that carry the $A$ population-structuring allele. Thus, the $A$ structuring allele will exhibit positive linkage disequilibrium with the more cooperative behaviour, and conversely the $a$ structuring allele will exhibit positive linkage disequilibrium with the more selfish behaviour. Since by definition cooperation increases mean fitness, and cooperation has become linked with the $A$ structural trait, individuals with the $A$ structural trait have the component of their fitness that is due to social behaviour (i.e., fitness affects from interactions with others) increased, on average, compared to individuals with the $a$ trait. Thus the $A$ structural allele, that supports cooperation, will increase in frequency, all other factors being equal. Our argument thus implies that selection on social behaviour can induce an indirect component of selection on population-structuring traits, through the generation of linkage disequilibrium. Moreover, the \emph{component} of selection on population structure which derives from selection on social behaviour must favour the creation of structures that support cooperation, rather than selfishness \citep{Powers:2010:a}. 

The above argument makes two critical but logical assumptions. The first of these is that individuals with structural allele $A$ find themselves living in structure $A$, whereas individuals with allele $a$ find themselves living in structure $a$. Essentially, what matters is that a structural differential exists, such that individuals with allele $A$ on average experience a different population structure, and hence potentially a different selective environment, to those with allele $a$. For example, if the $A$ allele coded for a greater degree of inbreeding, then individuals with the $A$ allele should inbreed with a greater probability than individuals without the allele (see \citealt{Breden:1991:a}). Likewise, if the allele codes for a group size preference, then individuals with a smaller group size preference allele should, on average, find themselves in smaller groups than individuals without the allele. If this were not the case, then linkage disequilibrium would not be generated between the socio-behavioural and population-structuring alleles, and so selection on population structure would not be induced by the above mechanism. The second key assumption is that the different population structures select for different levels of cooperation. If this were not the case, for example if both population structures selected for zero cooperation, then selection could not act on population structure via the social trait. 

In the model presented below, we consider the introduction of new population-structuring alleles by small mutations from existing ones. This model serves to illustrate the logical argument presented above. It also serves to elucidate what the assumptions of the argument mean for the specific case of group size evolution. In particular, it illustrates how the validity of the second assumption is affected by the type of social interaction (Prisoner's Dilemma or Snowdrift game type of cooperation; \citealt{Doebeli:2005:a}). Finally, the model demonstrates that a succession
of small mutations on group size can be selected and cause a population to evolve from an initial group size little
conducive to cooperation, to one highly conducive to cooperative behaviour.

\section*{The concurrent evolution of initial group size preference and public goods production} 

To illustrate the above argument, we consider the concurrent evolution of the number of individuals that found a group (which we hereafter refer to as founding size or simply group size) with public goods production. Public goods are those produced by an individual and shared with other group members \citep{Driscoll:2010:a}. They typically increase the fitness of all group members, but at a unilateral cost to the producer. Examples of such public goods production are widespread in nature and include the production of extra-cellular substances by microbes \citep{Griffin:2004:a,Gore:2009:a}, and the sharing of information by an individual with the rest of its group, as occurs during predator inspection by guppies \citep{Dugatkin:1990:a}, and alarm calls in flocks of birds \citep{Charnov:1975:a}. The production of such goods is a type of cooperative behaviour \citep{West:2007:b} that is vulnerable to exploitation by cheating non-producers, which reap the benefit of the public good without contributing to it. This can lead to a ``Tragedy of the Commons'' \citep{Hardin:1968:a,Rankin:2007:a}, in which cheating non-producers increase in frequency, even though this leads to a decline in mean fitness. 

However, the ``tragedy'' can potentially be averted in a group-structured population. In group-structured populations, individual selection on social behaviour (kin selection) can be partitioned into two components \citep{Price:1972:a,Hamilton:1975:a,Wilson:1975:a}, as follows. First, within each social group cooperators may decline in frequency due to exploitation by selfish cheats; this is within-group selection \citep{Wilson:1975:a}. Second, groups founded by different proportions of cooperative individuals may grow to different sizes by the time of dispersal, and hence contribute different numbers of individuals into the migrant pool; this is between-group selection \citep{Wilson:1975:a}. If groups with more cooperators contribute more individuals into the migrant pool, then cooperation may be able to evolve despite its local disadvantage within groups \citep{Wilson:1975:a,Wilson:1994:a}. 

The amount of cooperation that evolves in a particular case depends on the proportion of the total genetic variance, at the locus for public goods production, that is between groups \citep{Wilson:1975:a,Hamilton:1975:a} (or equivalently, the genetic relatedness of group members \citep{Queller:1992:a,Lehmann:2007:a}; see \textit{Discussion}). Any factor that increases the genetic variance between groups will favour greater cooperation. These include the founding group size, and the genealogical relatedness of the founding members. Here, we investigate how the size of social groups when they are formed co-evolves with public goods production. Specifically, if groups are formed randomly then between-group variance, and hence selection for cooperation, is inversely proportional to the founding size \citep{Wilson:1980:a}.  Founding group or colony size is influenced by genetic traits in many taxa, and can hence evolve by individual selection. Examples include queen number in social insects \citep{Ross:1991:a,Tsuji:1996:a}, and propagule size in colonial single-celled organisms such as choanoflagellates \citep{Michod:2000:a}, and bacteria in biofilms (see \textit{Discussion}). Indeed, the evolution of founding group size in colonial single-celled organisms has been argued to be a fundamental part of the transition to multicellularity \citep{Michod:2000:a,Roze:2001:a}.

\subsubsection*{The model}

Our model is based on the classic ``Haystack'', or aggregation and dispersal, model initially developed by \citet{Smith:1964:a} and later expanded into a general multi-level selection model (e.g., \citealt{Wilson:1981:a}, \citealt{Wilson:1987:a} and \citealt{Fletcher:2004:a}). We model a population of $N$ haploid asexually reproducing organisms, that periodically aggregate to form social groups in which the public good is shared. These groups stay together for $T$ generations, before all individuals synchronously disperse to form a global well-mixed migrant pool, from which new groups are formed (ecologically, dispersal and mixing might be triggered by local resource depletion, e.g. \citealt{Wilson:1981:a}, \citealt{Hochberg:2008:a}; synchronous dispersal is assumed for ease of analysis). This process of group formation, reproduction for $T$ generations, and dispersal repeats for a number of cycles, $D$. Crucially we allow the founding group size, and hence relatedness, to evolve by individual selection. Thus, individual selection may create large founding groups (hence with low relatedness) that select for selfish behaviour, or small founding groups (consequently with high relatedness) that select for cooperation.

Let $t$ be a counter for the generations within groups in a single aggregation and dispersal cycle; it thus ranges from 0 to $T$, and is reset with every new cycle. Then, let $n_t$ be the size of a group at generation $t$. We denote the founding size of a new group at the beginning of a cycle, $n_{t=0}$, by $z$. This is influenced by individual group size preferences, which we represent with a multi-allelic integer locus on an individual's genotype, $\gamma$. We assume that, to a first approximation, all individuals are able to form groups that satisfy their size preference, such that the number of groups of founding size $z$, $G_z$, is $N_{\gamma=z}/z$, where $N_{\gamma=z}$ is the number of individuals with group size preference allele $\gamma=z$ in the global population. An individual's genotype also contains a biallelic locus that codes for their social behaviour, i.e., whether they are cooperative and contribute to the public good, or are selfish and do not. Thus, each size preference allele is paired with a frequency of the cooperative allele (which can be different for different size preference alleles), which is used when determining the initial frequency of cooperation in each group. Formally, the expected count of groups of founding size $z$ (assuming that there are $z$ or more individuals with size preference allele $\gamma=z$ in the population, otherwise the count is zero) with $a$ cooperators is given by the function $h\left(z,a\right)$:

\begin{align}
h\left(z,a\right) &= G_z\left[\frac{\binom{A_{\gamma=z}}{a}\binom{N_{\gamma=z}-A_{\gamma=z}}{z-a}}{\binom{N_{\gamma=z}}{z}}\right],
\label{eqnNumGroups}
\end{align} 

where $G_z$ is the total count of groups of founding size $z$, calculated as described above, $A_{\gamma=z}$ is the number of cooperators with size preference allele $\gamma=z$, and $N_{\gamma=z}$ is the total number of individuals (cooperative and selfish) with size preference allele $\gamma=z$. The term in square brackets is the hypergeometric distribution of the cooperative allele to groups of founding size $z$, and represents random sampling of individuals with a given size preference allele without replacement, i.e., groups are not formed assortatively on social behaviour. Crucially, this distribution of social behaviours is taken only over individuals with size preference allele $\gamma=z$, rather than over the whole population. This means that if linkage disequilibrium develops between socio-behavioural and size preference alleles, then different size preference alleles can have different means and variances in the proportion of cooperators they experience,  As a control, the possibility of linkage disequilibrium developing can be removed by replacing $A_{\gamma=z}$ and $N_{\gamma=z}$ with $A$ and $N$, the count of cooperators over all size preference alleles, and the total population size, respectively.

The function $h\left(z,a\right)$ gives the distribution of individuals to groups at each group formation (aggregation) stage. After group formation, reproduction and selection occur within each group for $T$ generations. In each generation the number of cooperators ($a$) and selfish individuals ($s$) in a group, and hence the size of the group ($n$), change in a manner that is sensitive to the proportion of cooperators in that group, according to the following recursive equations:

\begin{align}
a_t &= a_{t-1}\Big(1+w_0+\rho_a\left(a_{t-1},n_{t-1}\right)\Big) \label{eqnWithinGroups1}\\
s_t &= s_{t-1}\Big(1+w_0+\rho_s\left(a_{t-1},n_{t-1}\right)\Big) \label{eqnWithinGroups2}\\
n_t &= a_t + s_t \label{eqnWithinGroups3}
\end{align}

where $\rho_a(a_{t-1},n_{t-1})$, defined below, is the fitness payoff that a cooperator in a group of size $n$ with $a-1$ other cooperators receives from social interactions, $\rho_s(a_{t-1},n_{t-1})$, also defined below, is the fitness payoff a selfish individual in the same group receives, and $w_0$ is a baseline fitness in the absence of social interactions. We allow $a$, $s$, and $n$ to take continuous values, such that offspring counts are not converted to an integer but are left as a real number (they are, however, rounded to an integer when applying Equation~\ref{eqnNumGroups} to generate the distribution of groups at the start of a new cycle).

Finally the number of cooperators in the global population with size preference allele $\gamma=z$, $A_{\gamma=z}$, at the end of an aggregation and dispersal cycle, $d$, (i.e., after group formation by Equation~\ref{eqnNumGroups}, and $T$ generations of reproduction and fitness-proportionate selection within groups given by $T$ iterations of equations~\ref{eqnWithinGroups1}-~\ref{eqnWithinGroups3}) is given by:

\begin{align} 
A_{\gamma=z} &= \sum_{i=0}^{z} h(z,i)a_T(i).
\label{eqnGlobalNumCoop}
\end{align}

This is the number of cooperators contributed to the global population by all possible groups of founding size $z$, multiplied by the expected count of that type of group (where the number of cooperators when the group is founded, $i$, varies from 0 to $z$, and $a_T(i)$ is the number of cooperators contributed by a group founded by $i$ cooperators and $z-i$ selfish individuals, after $T$ iterations of equations~\ref{eqnWithinGroups1}-~\ref{eqnWithinGroups3}). Likewise, the total number of selfish individuals with size preference allele $z$ after $T$ generations within groups is:

\begin{align}
S_{\gamma=z} &= \sum_{i=0}^{z} h(z,i)s_T(i).
\label{eqnGlobalNumSelfish}
\end{align}    
 
Our model consists of repeated iterations of equations~\ref{eqnGlobalNumCoop} and~\ref{eqnGlobalNumSelfish} (which depend on quantities from equations~\ref{eqnNumGroups}-~\ref{eqnWithinGroups3}), corresponding to repeated aggregation and dispersal cycles.

\subsubsection*{Within-group payoff functions}

We model the fitness payoffs that cooperative producers and selfish non-producers of the public good within a group receive using the $n$-player Prisoner's Dilemma and Snowdrift games \citep{Doebeli:2005:a}, where $n$ is the group size. The difference between the $n$-player Prisoner's Dilemma and Snowdrift games is whether selection on social behaviour is directional towards selfish behaviour at fixation, or is negative frequency-dependent leading to a polymorphism of behaviours. Our prior work suggests that these different types of selection on social behaviour within groups will have important effects on whether the second assumption of our logical argument holds, i.e., on whether small mutational differences in population structure will select for differing amounts of cooperation \citep{Powers:2008:a}, and hence allow positive linkage disequilibrium between cooperative and population-structuring alleles to develop. The $n$-player Prisoner's Dilemma represents the standard conception of altruism in evolutionary biology \citep{Fletcher:2007:a}, whereby selfish non-producers are always fitter than cooperative producers within their same group. This is despite the fact that group productivity increases with the proportion of cooperators, thereby creating a tragedy of the commons situation \citep{Hardin:1968:a,Hardin:1971:a,Rankin:2007:a} in which selection is directional towards all individuals being selfish non-producers. However, another scenario is that cooperative producers are fitter at low frequency, and less fit at high frequency, leading to a polymorphism of cooperative and selfish behaviours within a group. This can occur when cooperators are able to internalise some of the good that they produce, such that they receive a greater \textit{per capita} share of the benefits. However, if the cost remains fixed, but the benefit of the good tails off with increasing frequency, then above a threshold frequency selfish individuals will become fitter \citep{Hauert:2006:b}. This describes a negative frequency-dependent selection scenario, leading to a polymorphism of cooperative and selfish behaviours within a social group. 

The payoff matrix for the 2-player version of both the Prisoner's Dilemma and Snowdrift games is shown in Table~\ref{tabSD}. Under this payoff structure, cooperators provide a benefit $b$ to themselves and their social partner, at a cost to themselves of $c$. If their partner also cooperates, however, then the cost is shared ($c/2$). When $0.5<b/c<1$, this payoff structure produces the Prisoner's Dilemma, representing directional selection favouring selfish behaviour. Conversely, when $b/c>1$ the Snowdrift game, representing negative frequency-dependent selection, is produced\citep{Doebeli:2005:a}. The payoffs from pairwise interactions in Table~\ref{tabSD} are generalised to social groups of size $n$ by treating each group as a well-mixed population, such that pairs of individuals within the group interact at random. In this case, each individual experiences the \emph{proportion} of cooperators within its social group. We thus multiply the fitness payoffs for interacting with a certain type of individual (cooperative or selfish) with the proportion of that type within the focal individual's group (note that other generalisations of the Prisoner's Dilemma and Snowdrift games to $n$-players are possible that produce the same dynamics of directional or frequency-dependent selection within social groups, respectively; see, for example, \citealt{Hauert:2006:b}). Doing so yields the following payoff function for individuals within a social group, where $a/n$ is the proportion of cooperators ($p$) within the group:

\begin{table*}[htbp]
	\centering
	\caption{Fitness payoff matrix for the Snowdrift and Prisoner's Dilemma games \citep{Doebeli:2005:a}. This payoff structure produces directional selection for selfish behaviour within a group (the Prisoner's Dilemma) when $0.5<b/c<1$, and negative frequency-dependent selection leading to a polymorphism of cooperative and selfish behaviours (the Snowdrift game) when $b/c>1$.}
		\begin{tabular}{|l|l|l|}
		\hline
		& \textbf{Cooperate} & \textbf{Selfish} \\ \hline
		Payoff to Cooperate & $b-c/2$ & $b-c$ \\ \hline
		Payoff to Selfish & $b$ & 0 \\ \hline 
		\end{tabular}
	
	\label{tabSD}
\end{table*}

\begin{align}
\rho_a\left(a,n\right) &= \frac{a}{n}\left( {b - \frac{c}{2}} \right) + \left( {1 - \frac{a}{n}} \right)\left( {b - c} \right), \label{eqnMutModelSDcoop} \\
\rho_s\left(a,n\right) &= \frac{a}{n}b.
\label{eqnMutModelSDselfish}
\end{align}

\begin{figure}[htb]
\centering
\subfigure[]{\label{figPDFitness}\includegraphics[scale=0.15]{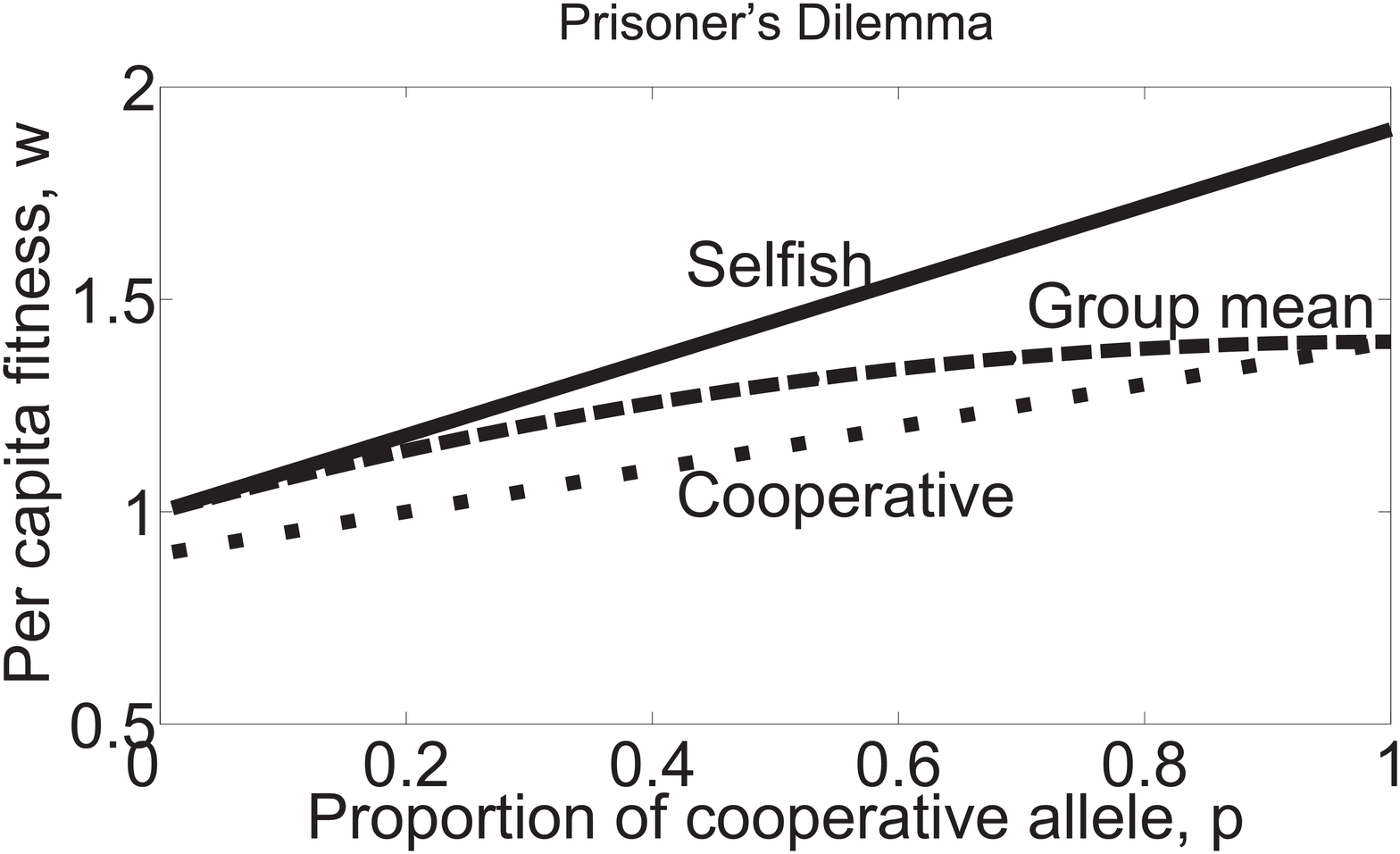}} 
\\ \subfigure[]{\label{figSDFitness}\includegraphics[scale=0.15]{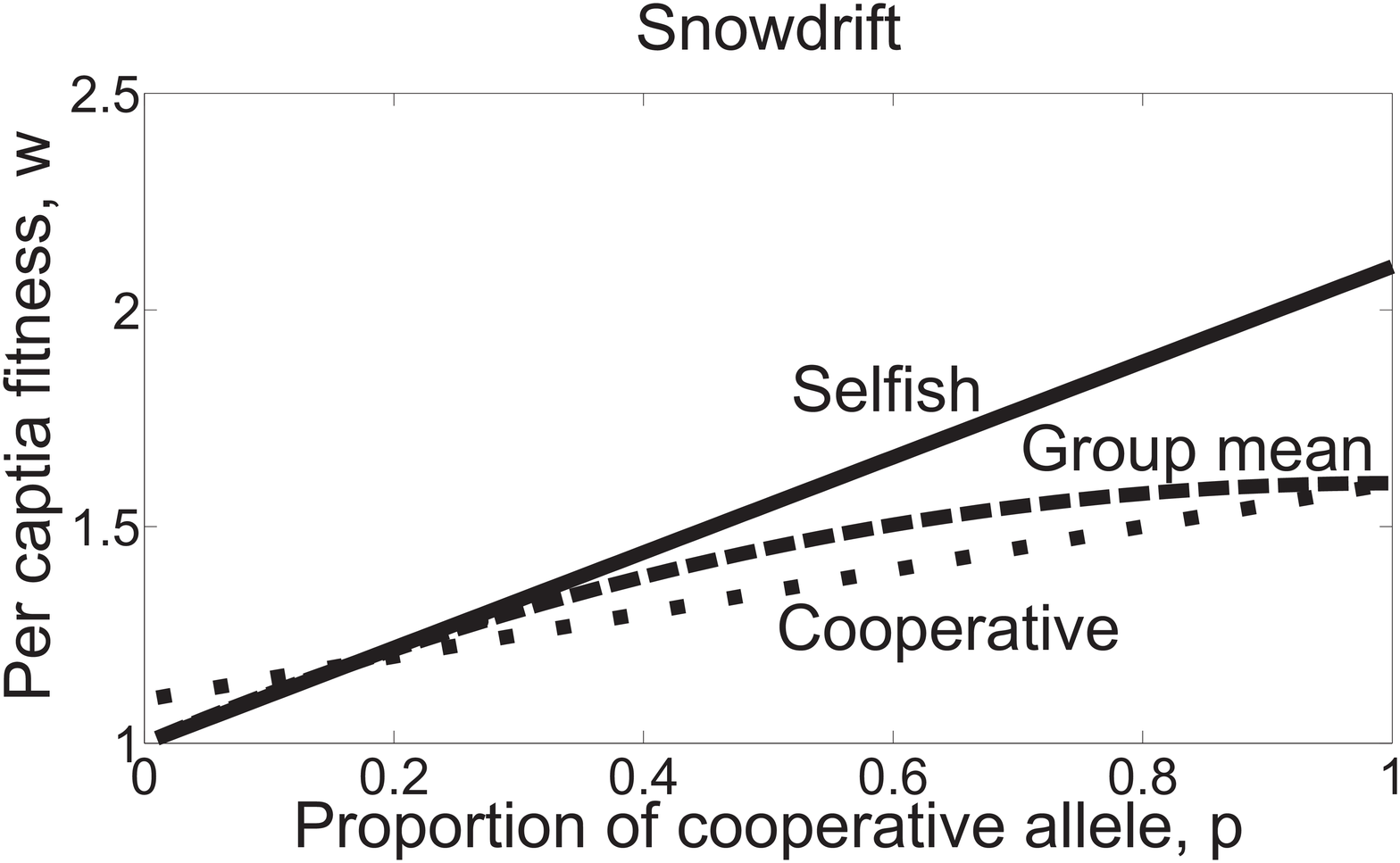}}
\caption{\textit{Per capita} fitness as a function of the frequency of the cooperative allele within the group in \textit{a}) Prisoner's Dilemma (i.e., $b/c=0.9$), \textit{b}) Snowdrift ($b/c=1.1$), games. A polymorphic within-group equilibrium exists in the Snowdrift, but not Prisoner's Dilemma, game. A polymorphic equilibrium is indicated graphically where the lines of the fitness functions cross, meaning that both types have equal fitness.}
\label{figFitnessFuncs}
\end{figure}

In both the Snowdrift and Prisoner's Dilemma parameterisations of these functions, individual selection within a group leads to an outcome that is suboptimal for the group, in terms of mean fitness of the group members. Specifically, although in both scenarios 100\% cooperate provides the highest mean fitness, in the Snowdrift game the equilibrium frequency of cooperators in an unstructured population is less than 100\%, and in the Prisoner's Dilemma case it is zero.  Thus, both games describe social dilemmas in which group and individual interests are not aligned. The plot of equations~\ref{eqnMutModelSDcoop} and~\ref{eqnMutModelSDselfish} in Figure~\ref{figFitnessFuncs} illustrates this graphically.

If these payoff functions were iterated within a single group until an equilibrium was reached, then the proportion of cooperators within that one group would be $1-c/(2b-c)$ \citep{Doebeli:2005:a}. This corresponds to the selfish allele at fixation within the group under the Prisoner's Dilemma parameterisation, and a stable polymorphism of cooperative and selfish alleles under the Snowdrift game parameterisation. Thus, selection on social behaviour is directional in the first case, and negative frequency-dependent in the second. Crucially, however, the social groups in our model stay together for only $t$ generations before dispersing and mixing. Hence, if $t$ is small then the equilibrium allele frequencies for a single group will not be reached. Rather, different groups may contain different proportions of cooperators at the dispersal stage. Because cooperation raises mean fitness (the greater the value of $a/n$, the greater the payoff to all group members in equations~\ref{eqnMutModelSDcoop}-~\ref{eqnMutModelSDselfish}; Figure~\ref{figFitnessFuncs}), groups with a greater proportion of cooperators will contribute more individuals into the global population at the dispersal stage. In this way, population structure can affect the \emph{global} equilibrium frequency of cooperation.  Dispersal before the within-group equilibrium frequencies are reached is necessary for such an effect to occur \citep{Wilson:1987:a,West:2002:a}, hence $T$ cannot be too large if population structure is to affect selection on social behaviour in this model. Existing theory predicts that the effect of population structure in this model should peak with an intermediate number of generations spent within groups \citep{Wilson:1981:a,Wilson:1987:a,Fletcher:2004:a}. However at least in the case of the Snowdrift game, which represents a kind of weak altruism where cooperation increases the \emph{absolute} fitness of the actor, we would still expect an effect of population structure when $T=1$ \citep{Wilson:1975:a,Wilson:1980:a}.   

Our use of game theoretic payoff functions does have the limitation that it assumes discrete cooperative and selfish phenotypes, rather than the more realistic case of the social phenotype representing a continuous degree of investment in the public good, or the probability that an individual contributes. On the other hand, the game theoretic payoff functions that we use provide a means to capture both directional and negative-frequency dependent selection through a simple parameterisation, thereby allowing for their comparison without changing any other aspect of the model.

\section*{Numerical analysis}

We examine the generation of linkage disequilibrium between socio-behavioural and population-structuring alleles, and the consequent selection for small founding group sizes that support cooperation, below. Closed form analysis for the generation of linkage disequilibrium in a social setting is known to be non-trivial \citep{Roze:2005:a,Hochberg:2008:a}, especially when one of the focal loci affects population structure and hence relatedness at the other locus \citep{Gardner:2007:a}. For example, the commonly used ``direct fitness'' method of \citet{Taylor:1996:a} does not describe the development of associations between loci \citep{Gardner:2007:a,Hochberg:2008:a}. Tracking the development of these associations, however, is fundamental for illustrating our logical argument. To obtain a closed form analytical solution that illustrates our argument, we would thus need a higher-order approximation that can explicitly account for changing group size and the generation of linkage disequilibrium. But at present, no such approximation is available. In particular, \citet{Gardner:2007:a} have recently provided a sophisticated analytical methodology for tracking the generation of linkage disequilibrium in a social setting, but acknowledge that further development is required to yield a general approach for tracking the evolution of relatedness. Thus, we analyse our model through numerical iteration of equations~\ref{eqnGlobalNumCoop} and~\ref{eqnGlobalNumSelfish} (see also \citealt{Peck:1988:a}, \citealt{Breden:1991:a}, \citealt{Aviles:2002:a} and \citealt{Hochberg:2008:a} for numerical and individual-based simulation treatments of the generation of linkage disequilibrium in a social setting). This allows us to explicitly track and account for the generation of linkage disequilibrium during the transients as well as at equilibrium, as required. Our results below show that the behaviour on the transients is fundamental to understanding the conditions under which population structure will evolve to support cooperation. 

Our approach is to consider a population initially fixed for a large group size preference allele, representing low between-group variance and hence little selection for cooperation, and then examine conditions under which mutant smaller group size alleles will be selected and increase between-group variance and selection for cooperation. To do so, we introduce a mutation operator into our model. Additional details are provided in Appendix 1.  

\subsubsection*{Parameter settings}

The parameter settings stated in Table~\ref{tabMutModParams} are used for the simulations presented below. We set $c=1$, with $b=0.9$ to yield Prisoner's Dilemma interactions, and $b=1.1$ to produce a Snowdrift scenario (this is close to the qualitative threshold between the two types of selection on social behaviour that occurs at $b/c=1$). In all of the simulations, we follow a population of total size $N=1000$, through 6000 group formation and dispersal cycles (preliminary experimentation revealed that this was a sufficient length of time for a global equilibrium of genotype frequencies to be reached, and the significant changes occur within the first 2000 cycles). We record the allelic and genotype frequencies averaged over 100 independent simulations. Group dispersal is set to occur after $t=5$ generations of selection and reproduction within groups. We start the population initially fixed for a size preference allele of 20, with the cooperative allele at an initial frequency of 0\% in the Prisoner's Dilemma case, and 16.67\% in the Snowdrift case (these are the equilibrium frequencies of the coopeerative allele in the simulations if group size is fixed at 20 and does not evolve, taken from the last 1000 aggregation and dispersal cycles and averaged over 100 runs). A fraction of $M=0.01$ individuals are mutated in the migrant pool between cycles. For those individuals chosen to undergo mutation, their size preference allele is mutated with a probability $m=0.9$, otherwise their socio-behavioural allele is mutated.  

\begin{table*}[htbp]
	\centering
	\caption{Model parameter settings.}
		\begin{tabular}{|p{4.5in}|l|}
		\hline
		\textbf{Parameter} & \textbf{Value} \\ \hline
		Cost to cooperating, $c$ & 1 \\ \hline
		Benefit to cooperating (Prisoner's Dilemma), $b$ & 0.9 \\ \hline
		Benefit to cooperating (Snowdrift), $b$ & 1.1 \\ \hline
		Fraction of population mutated, $M$ & 0.01 \\ \hline
		Probability any mutation is on size preference allele, $m$ & 0.9 \\ \hline
		Generations within groups before dispersal, $t$ & 5 \\ \hline
		Value of size allele fixed in initial population & 20 \\ \hline
		Smallest possible size allele & 1 \\ \hline
		Largest possible size allele & 40 \\ \hline
		Frequency of cooperative allele in migrant pool at initialisation (Prisoner's Dilemma) & 0 \\  \hline
		Frequency of cooperative allele in migrant pool at initialisation (Snowdrift) & 0.1667 \\ \hline 
		Gradient of sigmoidal fitness bonus function from Allee effect, $\mu$ & 0.4 \\ \hline
		Determinant of maximum benefit from Allee effect, $\beta$ & 1 \\ \hline
		Migrant pool density, $N$ & 1000 \\ \hline
		Number of aggregation and dispersal cycles, $T$ & 6000 \\ \hline  	
		\end{tabular}
	
	\label{tabMutModParams}
\end{table*} 

\subsubsection*{Directional selection on social behaviour represented by the Prisoner's Dilemma within groups}

\begin{figure}[!htb]
\centering
\subfigure[]{\label{figNoAlleePDSize}\includegraphics[scale=0.15]{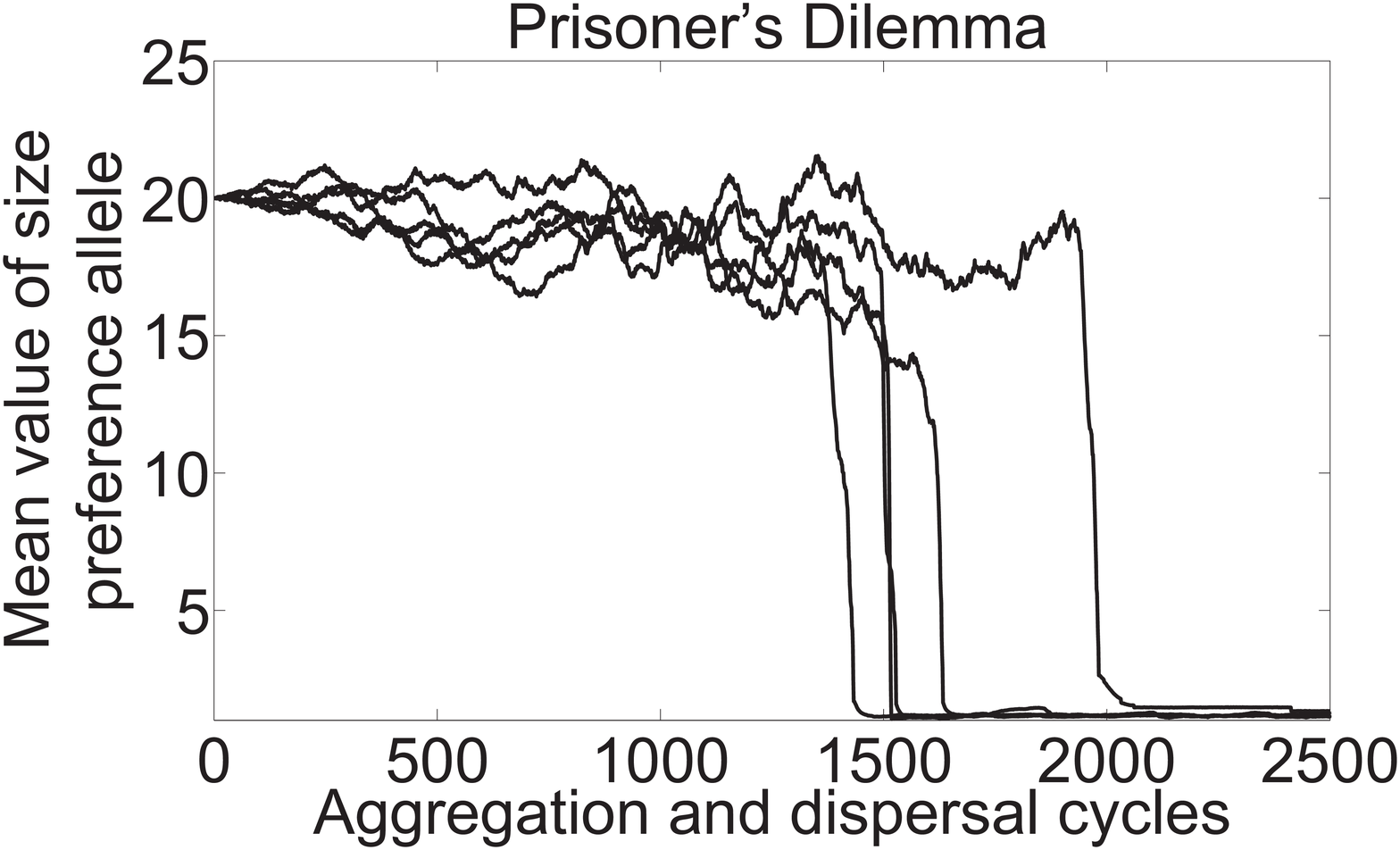}} 
\\ \subfigure[]{\label{figNoAlleeSDSize}\includegraphics[scale=0.15]{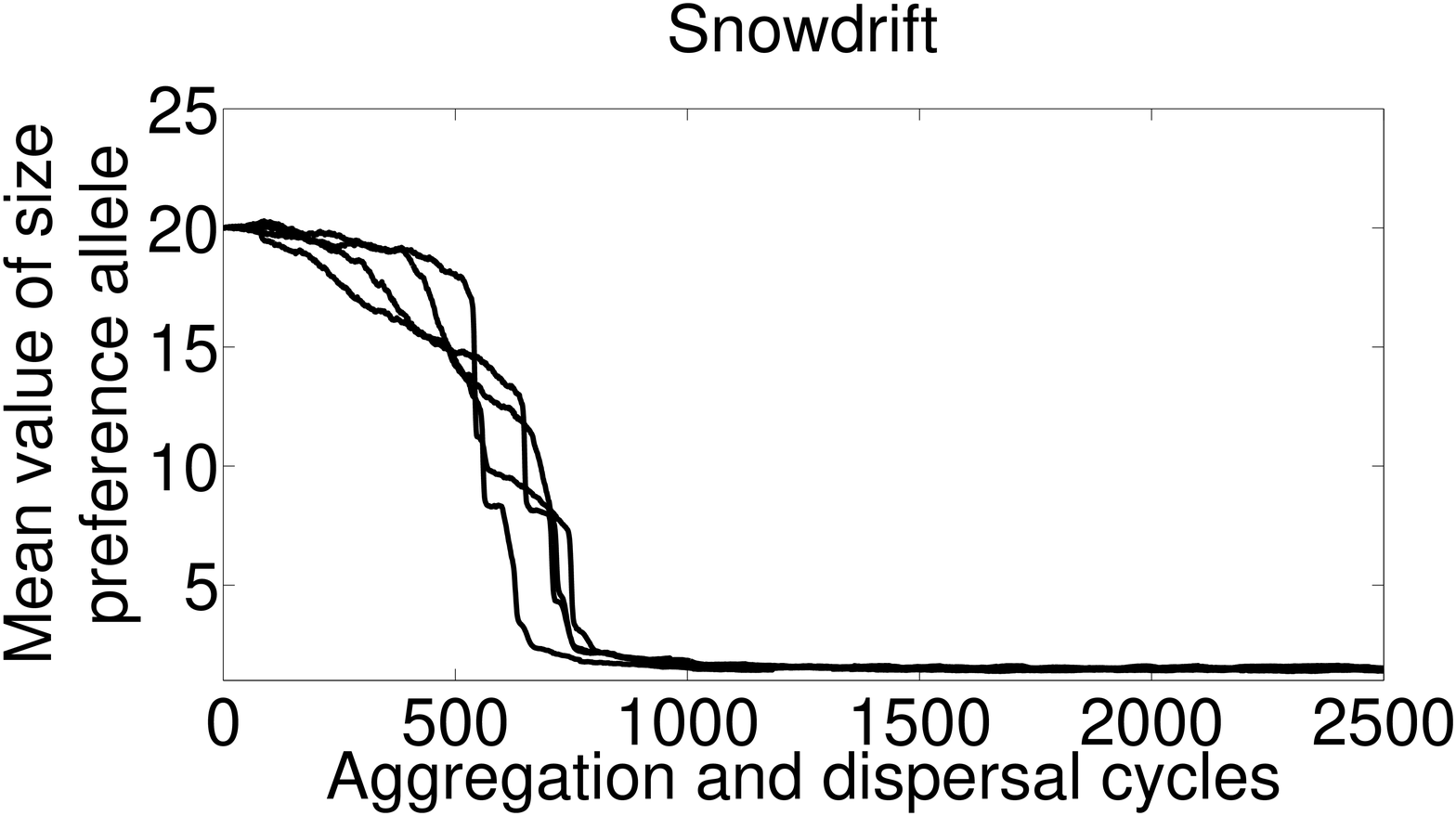}}
\\ \subfigure[]{\label{figNoAlleePropCoop}\includegraphics[scale=0.15]{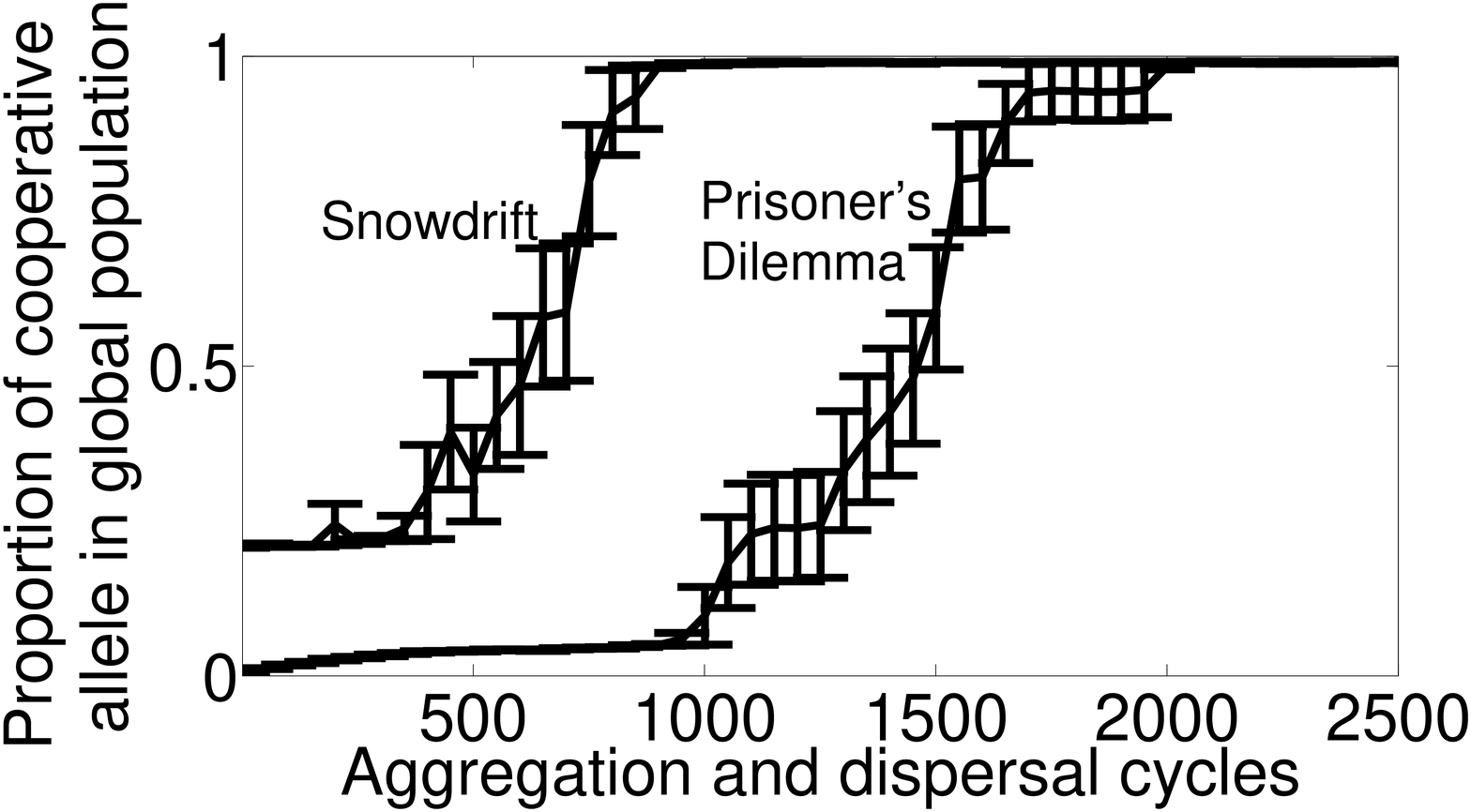}}
\
\caption[Concurrent evolution of founding group size preference and cooperation with no direct selection on group size.]{Concurrent evolution of founding group size preference and cooperation with no direct selection on group size. A) Mean value of size preference allele under Prisoner's Dilemma interactions (directional selection on social behaviour) in 5 independent runs. B) Mean value of size preference allele under Snowdrift interactions (negative frequency-dependent selection on social behaviour in 5 independent runs. C) Proportion of cooperative allele in global population, averaged over 30 runs (error bars show the standard error).      }
\label{figNoAlleeResults}
\end{figure}

We have investigated whether an individual adaptive gradient towards cooperative groups with high relatedness exists, and whether it can be followed when new group size preference alleles arise at mutation frequency. We first studied the case where selection on social behaviour is directional, representing a tragedy of the commons within groups, which we model using the $n$-player Prisoner's Dilemma as defined above. Our results show that, from an initial value of 20, the equilibrium mean size preference allele was close to the minimum possible (our results show numerically that this is an equilibrium, since recurring mutations occur on both group size preference and socio-behavioural alleles from this state, but do not cause a shift in the population mean) ,  (Figure~\ref{figNoAlleePDSize}), i.e., 1, with the cooperative allele reaching fixation apart from recurring mutations (Figure~\ref{figNoAlleePropCoop}). A founding group size of 1 maximises the relatedness of group members in subsequent generations (since all group members will be descendants of that one ancestor), and hence selects for maximal cooperation according to Hamilton's rule \citep{Hamilton:1964:a}. Thus, the equilibrium population structure is the one that selects for maximal cooperation. This result agrees with the logical argument introduced at the beginning of this paper. In particular, there is no other component of selection on the group size allele apart from that induced indirectly by selection on social behaviour. That is, group size only enters into equations~\ref{eqnWithinGroups1}-~\ref{eqnWithinGroups2} and~\ref{eqnMutModelSDcoop}-~\ref{eqnMutModelSDselfish} when calculating the proportion of cooperators within the group -- it has no other intrinsic effect and so cannot create a selection differential except through its effect on cooperation. Thus, because there is no other component of selection on the group size allele, a size allele which selects for, and hence exhibits positive linkage disequilibrium with, cooperation is favoured over one which does not.

To verify that selection for a smaller size allele would not occur without the generation of positive linkage disequilibrium with cooperation, we removed the possibility for sustained linkage disequilibrium. We did this by forming groups based on the size preference allele as before, but by drawing the behavioural alleles for a given group from the global frequencies, rather than using those of the participating genotypes. Thus, each group would centre around the global mean frequency of cooperation, rather than on the mean for its particular size. With this change to the model, there was no significant evolution of the group size allele, and the cooperative allele did not increase in frequency. Although cooperation is more frequent when averaged over small groups compared to large groups prior to the dispersal stage, this advantage is lost when the next generation of groups are formed, since any increase in cooperation is distributed across all group sizes uniformly. Thus, although individuals with smaller size alleles create the conditions for greater cooperation, they do not preferentially receive the benefits of this cooperation, and so do not have their fitness increased relative to individuals with larger size alleles. On the other hand, in the normal model linkage disequilibrium is generated between the alleles, and this allows group size alleles that create the conditions for cooperation to preferentially enjoy the consequent benefits.

\begin{figure}[h]
\centering
\subfigure[]{\label{figPDHeatMapNoAllee}\includegraphics[scale=0.17]{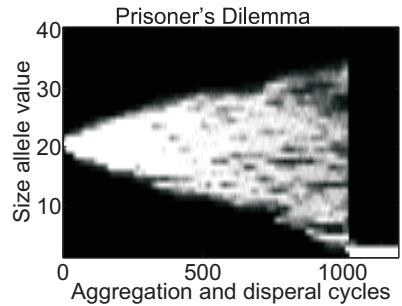}} 
\\ \subfigure[]{\label{figSDHeatMapNoAllee}\includegraphics[scale=0.17]{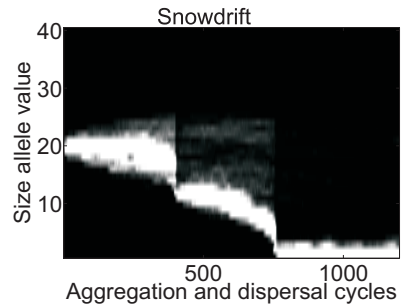}}
\caption[Group size allele evolution with no direct selection on group size.]{A) Size allele frequency evolution during a typical run under Prisoner's Dilemma interactions (lighter shades represent greater frequency of the allele value in the population). B) Size allele frequency evolution during a typical run under Snowdrift interactions.}
\end{figure}

Closer inspection of the results where linkage disequilibrium develops shows that the selection pressure favouring small groups is punctuated, not gradual. The plot of size allele frequencies over time in Figure~\ref{figPDHeatMapNoAllee} shows that in the initial stages there is no significant component of selection pressure on population structure. Specifically, the allele frequencies do not gradually shift downwards as mutations accumulate, instead, size preferences spread out equally in both directions until such a time as a preference for very small groups arises by mutation, at which point that very small size allele rapidly fixes in the population. This is shown by both Figure~\ref{figPDHeatMapNoAllee} and the step change in the mean value of the size allele in Figure~\ref{figNoAlleePDSize}, which illustrates that no significant selection occurs until a very small size preference is reached (we show independent representative runs rather than the mean in Figure~\ref{figNoAlleePDSize} to highlight this step change, which would be smoothed out when taking an average over multiple runs, since it happens at different times on different runs due to the stochasticity of the mutation process). The fact that the allele frequencies initially spread out in both directions shows that no significant adaptive gradient on this allele is present at the start. This implies that small, cooperative, groups might not evolve at all if the starting group size is much larger than the 20 we use. This is despite the fact that the mean fitness of individuals would be higher in such groups, due to the benefits of cooperation.

\begin{figure}[htbf]
\centering
\includegraphics[scale=0.17]{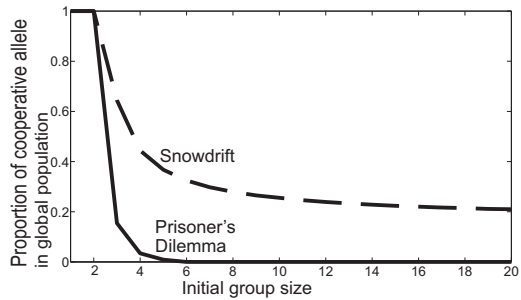} 
\caption{Global proportion of cooperative allele at equilibrium, for different (non-evolving) founding group sizes.  The solid line shows results where selection on social behaviour is directional towards selfishness within a group (the Prisoner's Dilemma). The dashed line shows results where selection on social behaviour within groups is instead negative frequency-dependent (the Snowdrift game).}
\label{figFixedGroupSize}
\end{figure}

Why, then, is there no significant selection pressure on founding group size in the initial stages? If a mutant smaller size preference is to be selectively advantageous, it must be the case that groups of that size have members which experience a greater frequency of cooperation than those of the current size. This follows immediately from equations~\ref{eqnWithinGroups1}-~\ref{eqnWithinGroups2} and~\ref{eqnMutModelSDcoop}-~\ref{eqnMutModelSDselfish}, since any differential fitness for bearers of the same behavioural allele must result from them experiencing a different frequency of cooperation (i.e., a different average value of $a/n$ in their groups); there are no other sources of differential fitness in equations~\ref{eqnWithinGroups1}-~\ref{eqnWithinGroups2} and~\ref{eqnMutModelSDcoop}-~\ref{eqnMutModelSDselfish}. So, for an adaptive gradient to exist, groups of founding size $z-1$ must enjoy more cooperation than those of size $z$ \citep{Powers:2008:a}. The solid line in Figure~\ref{figFixedGroupSize} explores this in the case where social interactions follow an $n$-player Prisoner's Dilemma, by calculating the frequency of cooperation in the global population over a range of non-evolving founding group sizes (taken over the last 1000 cycles, and averaged over 100 runs for each group size), with all other parameters the same as in the previous simulation (i.e., no assortment on behaviour during group formation, 5 generations within groups between dispersal episodes, and $b/c=0.9$).  This shows that, for $\gamma=z>6$, no adaptive gradient on the size allele can exist, since above this threshold moving to a slightly smaller group size does not increase cooperation, and so size allele $\gamma=z-1$ cannot, on average, be fitter than size allele $\gamma=z$. That is, the between-group component of selection on social behaviour is no greater when groups are of founding size $z-1$ than when they are of founding size $z$. This threshold for the start of an adaptive gradient can also be seen in Figure~\ref{figPDHeatMapNoAllee}, where allele frequencies only became concentrated around a few values once mutants with a size preference of 5 or less had arisen. These results suggest that an adaptive gradient towards the population structures favouring cooperation will only exist from a small range of initial conditions. However, we show in the next section that this is a consequence of assuming directional selection on social behaviour.

\subsubsection*{Negative frequency-dependent selection on social behaviour represented by the Snowdrift game within groups}

As previously discussed directional selection against cooperation, as modelled by the Prisoner's Dilemma, represents a worst-case scenario. Rather, the nature of cooperation may be such that selection is negative-frequency dependent, supporting a polymorphism of behaviours within groups. To investigate the effects of this kind of selection on the evolution of group size, we changed $b/c$ from 0.9 to 1.1 to produce the $n$-player Snowdrift game, and then reran the experiments in the previous section. The equilibrium mean initial group size is the same as under directional selection, that is, very close to 1, which in turn selects for maximal cooperation between group members during subsequent group growth. However, as figures~\ref{figNoAlleeSDSize} and \ref{figSDHeatMapNoAllee} show, the dynamics on the transient to reach this equilibrium are very different. Rather than the size preference allele spreading out in both directions, the mass of the allele frequencies show a trend of moving downwards from the start, suggesting the presence of an adaptive gradient towards a population structure that supports cooperation from the outset. 

To confirm this, we again considered whether a mutation from group size preference $\gamma=z$ to $\gamma=z-1$ would increase the amount of cooperation its bearers experienced. The dashed line in Figure~\ref{figFixedGroupSize} shows that if all groups are of (founding) size $z=20$, then decreasing group size by 1 always yields some increase in cooperation. The question is then: why does this occur under Snowdrift, but not Prisoner's Dilemma, interactions? We have shown elsewhere \citep{Powers:2008:a,Powers:2010:a} that negative frequency-dependent selection on social behaviour maintains some between-group variance over a much larger range of conditions than if selection is directional. This is because between-group variance is proportional to the frequency of the least common behavioural allele in the population (if social behaviours are distributed binomially when groups are formed then the between-group variance at the first generation is $P(1-P)/z$; \citealt{Wilson:1980:a}), and while this tends to zero under directional selection, it does not do so under negative frequency-dependence \citep{Powers:2008:a,Powers:2010:a}. This preservation of between-group variance means that some between-group component of selection on social behaviour can be seen over a much larger range of conditions and, consequently, moving to a slightly smaller initial group size increases cooperation over a much larger range.

The results presented so far are from a model in which there is no direct selection on the group size preference allele. This has allowed us to show that cooperation can exert indirect selection pressure on group size preference, driving its evolution. However, we might expect there to be other direct components of selection on group size preference. In the next section, we introduce a revised model in which there is such direction selection on group size. Moreover, the direction of this direct selection is in opposition to the indirect selection from cooperation, thereby making the evolution of a group size that supports cooperation more challenging.

\subsubsection*{Analysis with an opposing component of selection on population structure due to an Allee effect}

We introduce here a direct component of selection into the model that favours larger groups, representing a (weak) Allee effect \citep{Allee:1938:a,Odum:1954:a,Aviles:1999:a}. A larger founding group size may be favoured by, for example, better defence against predators, or access to resources that a smaller group cannot obtain \citep{Aviles:1999:a}. Such factors create a component of selection pressure on population structure that is in a direction opposite to that from social traits (i.e., they tend to reduce relatedness by creating pressure for a larger founding size). Consequently, this makes the evolution of population structures that support cooperation more difficult. 

This component of selection is represented by the following, positive density-dependent, function of group size, $\sigma(n)$:

\begin{align}
\sigma(n)  &= \frac{\beta}{{1 + e^{ - \mu n} }} - \frac{\beta}{2}. 
\label{eqnAllee3}
\end{align}

$\sigma(n)$ is a sigmoidal function of group size, with gradient $\mu$ (which determines how quickly the benefit tails off as the group grows), and $\beta$ a parameter which determines the maximum benefit. A sigmoidal function of group size is used to model the fact that, above a certain size, the advantages of number become cancelled out by the effects of increased crowding \citep{Odum:1954:a,Aviles:1999:a}. This direct selection on group size is incorporated into the model by replacing equations~\ref{eqnWithinGroups1}-~\ref{eqnWithinGroups2} with the following:

\begin{align}
a_t &= a_{t-1}\Big(1+w_0+\rho_a\left(a_{t-1},n_{t-1}+\sigma(n)\right)\Big), \label{eqnAllee1} \\
s_t &= s_{t-1}\Big(1+w_0+\rho_s\left(a_{t-1},n_{t-1}+\sigma(n)\right)\Big), \label{eqnAllee2}
\end{align}

To obtain the numerical results presented below, we set $\mu=0.4$ and $\beta=1$.  In the absence of any possibility for cooperation, a founding group size of 20 or greater would be favoured using these parameter settings for the sigmoidal function, since this is the group size for which the gradient reaches zero. However, given that group size can affect selection for cooperation, individual fitness would actually be increased with a smaller group size that produces more between-group variance, and hence selects for greater cooperation. We investigate below whether an adaptive gradient towards such an intermediate group size exists. 

\begin{figure}[!htb]
\centering
\subfigure[]{\label{figAlleeSizeFreq}\includegraphics[scale=0.15]{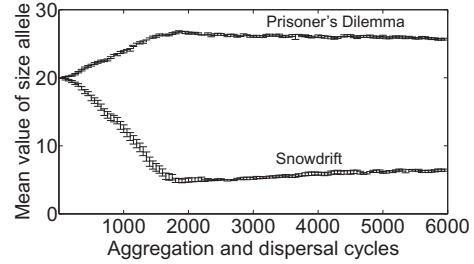}} 
\\ \subfigure[]{\label{figAlleePropCoop}\includegraphics[scale=0.15]{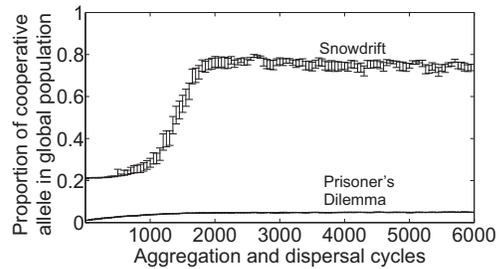}}
\\ \subfigure[]{\label{figBGVar}\includegraphics[scale=0.15]{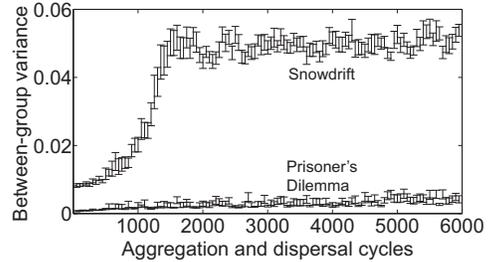}}
\\ \caption[Concurrent evolution of founding group size preference and cooperation with a direct opposing component of selection on population structure provided by an Allee effect.]{Concurrent evolution of founding group size preference and cooperation with a direct opposing component of selection on population structure provided by an Allee effect. Error bars show standard error from 100 independent trials. A) Mean value of size allele. B) Cooperative allele frequency in population. C) Evolution of between-group variance.  }
\label{figAlleeResults}
\end{figure}

\begin{figure}[!htb]
\centering
\subfigure[]{\label{figAlleePDHeatMap}\includegraphics[scale=0.17]{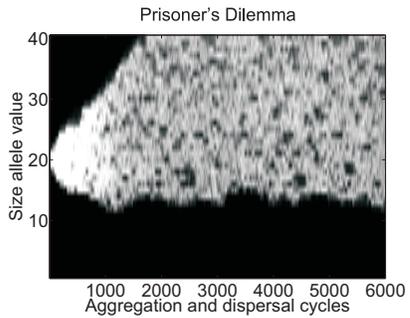}}
\\ \subfigure[]{\label{figAlleeSDHeatMap}\includegraphics[scale=0.17]{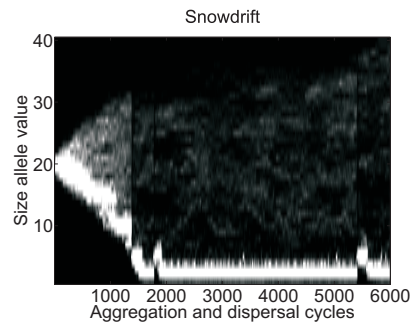}}
\caption[Group size allele evolution with opposing, direct, selection on group size from an Allee effect.]{Size allele frequency evolution (lighter shades represent greater frequency of the allele value in the population). A) Representative run under Prisoner's Dilemma interactions. B) Representative run under Snowdrift interactions. }
\end{figure}

The results in Figure~\ref{figAlleeResults} show how the evolution of founding group size is affected by the incorporation of an Allee effect. Founding group size preference evolves downwards, and between-group variance and hence cooperation increase, under Snowdrift but not Prisoner's Dilemma types of interaction (figures ~\ref{figAlleeSizeFreq}, ~\ref{figAlleePropCoop} and ~\ref{figBGVar}). This is because under Prisoner's Dilemma interactions, no selection on the size preference allele due to social behaviour exists from the initial state (Figure~\ref{figAlleePDHeatMap}), in accordance with our previous results. Because of the opposing selective force towards larger groups generated by the Allee effect, the size preference allele is no longer able to drift downwards. On the other hand, under Snowdrift interactions an adaptive gradient towards a smaller size preference still exists (Figure~\ref{figAlleeSDHeatMap}). Thus, these results highlight the importance of a small change in population structure causing an increase in cooperation. Without this, the evolution of population structures that support cooperation must rely on drift, which cannot overcome any opposing component of selection.

Interestingly, Figure~\ref{figAlleeSizeFreq} suggests that the mean value of the size preference allele starts to increase again under Snowdrift interactions once the cooperative allele has reached a high frequency. This is because at the end of a run linkage disequilibrium exists between the size preference and behavioural alleles, such that the selfish allele is associated with a large size preference, and the cooperative allele with a small preference, as predicted by our logical argument. Specifically, the selfish allele can only be maintained in large groups with little between-group variance. It is this association of the selfish allele with a larger size preference that pushes the mean value of the size allele slightly upwards towards the end of a run. We calculated the linkage disequilibrium at the end of a run (using Lewontin and Kojima's \citeyearpar{Lewontin:1960:a} normalised method), averaged over 100 trials, to be 0.62 (by contrast under Prisoner's Dilemma interactions the linkage disequilibrium was 0.12).  

\subsubsection*{Sensitivity analysis}

We have illustrated how a trait affecting the population structure of its bearers can evolve concurrently with social behaviour, and how a sharp qualitative distinction arises between Prisoner's Dilemma and Snowdrift style social interactions. In particular, the distinction is over whether a small change in population structure induces greater cooperation, and hence is selectively advantageous. The quantitative range in which this is true depends upon the area of parameter space over which a small change in group size increases between-group variance. In the model presented here, the other factors apart from founding group size that determine this are the initial state of the population (size preference and cooperation allele frequencies), the degree of assortativity on behaviour during group formation, the number of generations spent within groups before dispersal, and the cost-to-benefit ratio of cooperating. All of these factors matter, however, only in so far as they change the quantitative range over which small mutations on group size induce greater cooperation; the logical argument presented here still holds for other settings of these parameters. In a similar manner, the parameters of the Allee effect in Equation~\ref{eqnAllee3} determine the trade-off between the advantages of smaller initial size arising from selection for greater cooperation, versus the raw benefits of living with more individuals. However, while the exact position of this trade-off determines the final size preference that will evolve, it does not affect the qualitative result that the opportunity for cooperation can drive the evolution of population structure, nor the distinction between Prisoner's Dilemma and Snowdrift style social interactions. We thus do not wish to make claims with respect to the quantitative results of our simulations. The purpose of our model is not to make quantitative predictions, but to illustrate our logical argument and examine qualitative aspects of this process.

\section*{Discussion}

At the beginning of this article, we argued that cooperation could drive the evolution of population structure. The model presented above provides a simple illustration of this argument, for the particular population-structuring trait of founding group size preference. The model illustrates how a genetic preference for a smaller founding size can evolve because it increases between-group variance, and hence increases the benefits of cooperation that its bearers experience. In particular, the cooperative allele is selected for more strongly in smaller groups than in larger groups. Because the individuals in smaller groups tend to be those that have a genetic preference for such groups, the cooperative allele becomes linked, by selection, to small group size preference alleles. Since cooperation by definition raises mean fitness, the mean fitness of individuals with both the small group size and cooperation alleles is greater than the population mean fitness, so these genotypes increase in frequency.

This argument hinges on the development of linkage disequilibrium between small size preference and cooperative alleles. For this to occur by selection, smaller groups must select for a greater degree of cooperation, since the only source of differential fitness in the model (with no Allee effect) is that caused by experiencing different amounts of the benefits of cooperation (different mean $p$ values within groups). This is an instance of the general assumption of our logical argument, that different population structures must induce selection for different amounts of cooperation. Our results illustrate that whether selection on social behaviour within groups is directional or negative frequency-dependent is an important determinant of whether this assumption is likely to hold true. This is because directional selection, modelled here by the $n$-player Prisoner's Dilemma, necessarily limits the range of conditions over which some between-group variance can be generated \citep{Powers:2008:a,Powers:2010:a}. Consequently, this limits the range over which a small change in population structure can increase between-group variance and hence select for greater cooperation.

There is a growing interest in whether selection for cooperation is directional or negative frequency-dependent in a range of biological scenarios \citep{Doebeli:2005:a}. \citet{Gore:2009:a} have recently verified empirically that the public goods scenario of extra-cellular enzyme production in yeast does indeed follow a Snowdrift game. In the context of bacterial biofilms, \citet{Dugatkin:2003:a,Dugatkin:2005:a,Dugatkin:2005:b} have presented theoretical and empirical work arguing that antibiotic resistance is a public good that similarly undergoes negative frequency-dependent rather than directional selection, and hence follows a Snowdrift rather than Prisoner's Dilemma game. Similarly, \citet{Burmolle:2006:a} suggest that antibiotic resistance is a synergistic cooperative trait between multiple species, which again show a local coexistence of multiple types, rather than the competitive exclusion of the Prisoner's Dilemma. In the arthropod phylum, it has been shown that the success of social spiders who cheat during foraging is locally limited, suggesting negative frequency-dependent selection. In the case of mammals, \citet{Doebeli:2005:a} suggest that collective hunting in lions and baboons, and sentinel behaviour in meerkats, may for example also fit the $n$-player Snowdrift game. Our results imply that the evolution of population structure towards that which supports cooperation is much more plausible in these kinds of systems, as opposed to cases of strong altruism \citep{Wilson:1980:a} represented by the Prisoner's Dilemma. We suggest, however, that whilst cooperative traits which undergo negative frequency-dependent selection could initially drive the evolution of population structure, this evolved population structure could then provide a high enough relatedness to support the evolution of strongly altruistic traits. Thus, weakly altruistic traits subject to negative frequency-dependent selection (the Snowdrift game) could, by inducing selection on population structure, scaffold the subsequent evolution of strong altruism (the Prisoner's Dilemma). Through this mechanism, such traits may play an important role in the origin of high levels of sociality \citep{Powers:2010:a}.  

Our results can be understood in either a kin or group selection framework. In particular, the metric of between-group variance is equivalent to that of genetic relatedness, since relatedness is a measure of the correlation in behaviour between individuals within a social group compared to individuals chosen from the global population \citep{Grafen:1984:a,Queller:1985:a,Queller:1992:a}. We have shown here how selection on traits affecting population structure can lead to an increased relatedness between interacting individuals, and hence how a population structure that supports cooperation according to Hamilton's rule \citep{Hamilton:1964:a} can evolve. Thus, whilst Hamilton's rule predicts the conditions for cooperation to evolve, we have shown here how a population structure that satisfies those conditions can arise by adaptive evolution.

The joint evolution of population structure and social behaviour has also been considered in the context of the origin of multicellularity. \citet{Roze:2001:a} showed that when selfish mutations occur during the growth of clonal colonies (representing proto-multicellular organisms), it can be selectively advantageous for the group of cells to reproduce by breaking off single cells, rather than larger propagules of multiple cells. The reason for this is the same as in our model: a smaller initial group size increases between-group variance and hence reduces selection for selfish behaviour. We argue that this is in fact a general trend in the evolution of population structure, and our numerical model explicitly tracks the generation of linkage disequilibrium between group size and social behaviour when new initial group sizes arise by mutation. \citet{Pfeiffer:2003:a} used an individual-based simulation to model the co-evolution of clustering and cooperation between single cells, as a pathway to the origin of multicellular organisms. Also in the context of the transition to multicellularity, \citet{Hochberg:2008:a} considered the concurrent evolution of dispersal and social behaviour and illustrated the development of linkage disequilibrium between social and population-structuring traits. Specifically, they showed conditions under which selfish behaviour would become linked with a greater tendency to disperse from the cell group. Such work adds support to our argument that the generation of linkage disequilibrium between socio-behavioural and population-structuring traits is a fundamental force in the evolution of population structure, and should be explicitly considered when explaining the origin of sociality \citep{Powers:2010:a}.

We stress again that our logical argument does not only apply to founding group size. Rather, it applies to any heritable trait that modifies population structure so as to cause a difference in selection pressure on social behaviour between bearers and non-bearers. Our model was thus designed to provide a simple illustration of this argument, rather than to apply to any one system in particular. Nevertheless, it is worth fleshing out how our specific model of the evolution of founding group size might apply in principle to different social systems. There is recent empirical evidence that high relatedness, resulting from population structure, is important for the evolution of sociality in vertebrates \citep{Cornwallis:2010:a}. There is also evidence that some mammals can actively control the size of their group, by preventing migrants or solitaires from joining once the group is larger than their preferred size \citep[Chapter 4]{Giraldeau:2000:a}. Presumably such traits for group size regulation are at least partly genetic, and hence could co-evolve with cooperative behaviour in a manner analogous to our model. Indeed, the existence of a heritable group size preference has been verified empirically in birds \citep{Brown:2000:a}. The degree to which individuals are able to successfully regulate the size of their group will affect the degree to which the first assumption of our logical argument is met, i.e., the degree to which individuals with a heritable preference for a particular population structure are actually able to live in that structure. As this assumption is relaxed, for example, due to a cost of group size regulation, we would expect the generation of less linkage disequilibrium between socio-behavioural and population-structuring traits, and hence weaker selection towards the population structures that support cooperation.

Other social systems in which the evolution of founding group size might naturally apply include those where groups reproduce by propagule or fissioning. In these systems, which include social spider colonies \citep{Aviles:1993:a}, and bacterial micro-colonies within biofilms \citep{Hall-Stoodley:2004:a}, a new group is founded by a small sample of individuals from a single parent colony. The size of these founding propagules could potentially be affected by individual genetic traits, for example, by the amount of extra-cellular matrix secreted by bacteria in a biofilm. Thus, propagule size could co-evolve with social behaviour. In the next section, we discuss in detail how our theory of public goods production driving the evolution of founding group size can be tested empirically in bacterial biofilms.

\subsubsection*{Empirical hypotheses testable in bacterial biofilms}

Biofilms are formed when bacteria attach to a surface or interface and form large aggregate structures bound together by a co-produced extra-cellular matrix. They, rather than individual motile cells, are the most common mode of bacterial growth \citep{Ghannoum:2004:a}. They often exhibit a marked group structure, whereby bacteria within the biofilm tend to live in discrete micro-colonies \citep{Hall-Stoodley:2004:a} and moreover, produce and share various extra-cellular public goods \citep{Crespi:2001:a,West:2007:c}. There has recently been much interest in bacterial social evolution \citep{Rainey:2003:a,Griffin:2004:a,Kreft:2004:a,Buckling:2007:a}, but crucially previous theory assumes that population structure remains static.

Bacteria in biofilms exhibit a periodic dispersal cycle \citep{Hall-Stoodley:2004:a}, which is reminiscent of the ``Haystack'' aggregation and dispersal population structure that we have modelled. Micro-colonies form, grow, and disperse to form new colonies either via the shearing off of propagules containing varying numbers of individuals, or by the dispersal of single cells \citep{Hall-Stoodley:2004:a}. The structure of bacterial biofilm is generated by a complex interaction between ecological, environmental and genetic factors \citep{Ghannoum:2004:a,Flemming:2007:a,Xavier:2007:a}. However, as the production and nature of the extra-cellular polymers are genetically mediated, there is a strong possibility that dispersal mode and cooperation could co-evolve. In particular, we might expect that the factors affecting the balance between single cell and propagule dispersal, and mechanical and structural features of the extra-cellular matrix which affect propagule size, might be subject to the effect which we have described.

Micro-colony size and dispersal will be affected by a number of factors including shear stress, resource availability and competition. However, we predict that the creation of specific biofilm structure is at least partially driven by the benefits of cooperation. For example, the model suggests that smaller propagule size or increased dispersal of individual cells might evolve in response to the need for public goods production to be supported. This hypothesis could potentially be tested empirically by manipulating the need for a particular public good, such as siderophore production (see also \citealt{Brockhurst:2010:a} for complementary theory of how resource availability may change the cost and benefits of cooperation in biofilms). Siderophores are extracellular iron-scavenging molecules that are costly to produce, in terms of time and energy, but which aid bacterial metabolism in stressful low iron environments \citep{Ajit:2007:a}. Non-siderophore producing cheats arise \textit{in vivo} and would be assumed to prefer unstructured populations. If the need for siderophores is removed, for example by providing a saturation of iron, then average propagule size or propagule/single cell dispersal ratio may increase over numerous dispersal/recolonisation cycles due to the removal of the benefits of cooperation. This could occur either via drift or the upward pressure on size from any Allee effect. Indeed, a recent investigation into the evolution of extra-cellular matrix production \citep{Xavier:2007:a} implies the presence of such a selective force favouring greater matrix production, and hence potentially greater propagule size or reduced single-cell dispersal.

If iron is then made limiting once again via the addition of iron-chelating proteins, then propagule size should evolve back downwards, since siderophore production would again become beneficial, and more of this benefit would be bestowed upon individuals with traits leading to smaller propagule sizes and increased between-micro-colony variance. Thus, siderophore production should become linked with such traits, driving selection for smaller propagules. It should be noted that although we might expect the morphology of the biofilm to change in response to such differences in resource availability simply via ecological effects, we should be able to distinguish between these effects and evolutionary change. Rapid ecological responses should reverse quickly when we revert back to the original experimental scenario. We would expect that genetic change influencing population structure would display an increased latency when switching between regimes.

\section*{Conclusions}

We argue that considering the concurrent evolution of population structure and social behaviour in general provides a fundamental new perspective on the evolution of cooperation. In previous theories, where the population structure is static, cooperation is simply the adaptation of organisms' social behaviour to their current social environment. However, we argue that to explain the evolutionary origin of cooperation, it is necessary to explain why organisms live in that particular social environment, rather than in one that does not support cooperative behaviour. In this article, we have explicitly considered the joint equilibrium of population structure and social behaviour, and have shown how and why organisms move from living in a social environment little conducive to cooperation, to one where cooperative behaviour predominates. Thus, we can provide an explanation for the origin of a selective environment that supports cooperation, from whence, the evolution of cooperation itself is straightforward.

\section*{Acknowledgements}
We thank Stuart West, Claire El Mouden, and an anonymous reviewer for constructive suggestions. We also thank Seth Bullock and Rob Mills for helpful discussions, and Joel Parker for comments on an earlier version of the manuscript. The first author acknowledges funding from an EPSRC PhD Plus award at the University of Southampton.

\bibliographystyle{amnatnat}
\bibliography{lit}

\section*{Appendix 1: Simulation procedure}

Our numerical analysis of the model consists of a combination of individual-based and numerical simulation. Specifically, we assume that the migrant pool contains a finite number of discrete individuals. We model haploid genotypes with two loci, which reproduce asexually (i.e., with no recombination between loci). The first locus is biallelic and codes for whether an individual produces a public good at some unilateral cost. The second locus codes for an initial group size preference. A multi-allelic integer representation is used for this locus, which in the analysis presented here can take values between 1 and 40. Groups are formed by the following individual-based procedure, which approximates Equation~\ref{eqnNumGroups} while allowing for a finite number of individuals:

\begin{enumerate}
	\item Create a list of all individuals in the migrant pool.
	\item Sort this list in reverse order of group size preference, such that the individuals with the largest size preference are at the front of the list. Within each sub-list of individuals with the same size preference, randomise their position in the list with respect to their social behaviour (cooperative or selfish). 
	\item Create a new group, and add the individual at the front of the list to this group. Remove the added individual from the list.
	\item Continue adding individuals in order from the list, while the following condition is met: the mean size preference of the group members is less than the current group size. When this condition does not hold, advance to step 5.
	\item If there are still individuals in the list, go back to step 3, else all groups have been formed. 
\end{enumerate}

Regarding step 2 in this algorithm, randomising the order of each sub-list of individuals with the same size allele means that the behaviours are assigned to groups according to a hypergeometric distribution, and not assortatively. More precisely, there is a separate hypergeometric distribution of social behaviour for each value of the group size allele. Sorting the list in reverse size order handles the special case of the last group. This is because the last group will contain the handful of remaining individuals in the migrant pool. If the list was sorted in increasing order, then this last group would be small but would contain the individuals with the largest size preference. Sorting in reverse order means that the small last group contains the individuals with the smallest size preference. 

Once the groups have been formed, equations~\ref{eqnWithinGroups1}-~\ref{eqnWithinGroups2} are iterated recursively for $T$ timesteps (when calculating this recursion, a fractional number of individuals are allowed within groups). Note that by the above group formation procedure, a group can contain different size preference alleles. However, there is no selection on the group size allele \emph{within} groups, and hence the frequency of the size allele within a group changes only as a result of selection on social behaviour from equations~\ref{eqnWithinGroups1}-~\ref{eqnWithinGroups2}.

For computational convenience, we introduce a global population carrying capacity. Thus the total population size, $N$, remains fixed. This is achieved by multiplying $a_t$ and $s_t$, after each iteration of equations~\ref{eqnWithinGroups1}-~\ref{eqnWithinGroups2}, by a factor $N/N*$, where $N*$ is the new total population size after equations~\ref{eqnWithinGroups1}-~\ref{eqnWithinGroups2} have been iterated once for each group (as in \citealt{Fletcher:2004:a}). This scales group sizes back in a proportionate manner, leaving the proportions of genotypes within each group the same.  

After $T$ iterations of equations~\ref{eqnWithinGroups1}-~\ref{eqnWithinGroups2}, all groups disperse into a new global migrant pool (at which point the number of individuals is rounded to the nearest integer), and mutation occurs as follows.

A fraction $M$ of individuals in the migrant pool are randomly chosen to be mutated. Of this subset of the population chosen for mutation, a fraction $m$ have their size preference allele mutated, the remaining $1-m$ fraction their behavioural allele mutated; only one locus is mutated per individual, to illustrate that simultaneous mutations are not required to decrease group size and increase cooperation.  Mutation on the integer size preference allele occurs by decreasing its value by 1, with 50\% probability, or otherwise it is increased by 1. If the size allele is already at the upper or lower limit (40 or 1 with the parmater settings used here, respectively), then it always has 1 subtracted or added, respectively, if selected for mutation. Mutation on the biallelic behavioural allele (cooperative or selfish) occurs by switching to the other behaviour.

After mutation has occurred, the next generation of groups is formed from the migrant pool, and the aggregation and dispersal process is repeated for a sufficient number of cycles for an equilibrium to be reached.

\end{document}